\documentclass[pre,showpacs,amsfonts,amssymb,eqsecnum,%
twocolumn,floatfix]{revtex4}
\pdfoutput=1
\usepackage{amsmath}
\usepackage{graphicx}
\usepackage{bm}
\usepackage{color}
\allowdisplaybreaks

\begin{document}

\title{Transport properties of random walks under stochastic non-instantaneous resetting}
	
	\author{Axel Mas{\'o}-Puigdellosas}
	\affiliation{Grup de F{\'i}sica Estad\'{i}stica.  Departament de F{\'i}sica.
		Facultat de Ci\`encies. Edifici Cc. Universitat Aut\`{o}noma de Barcelona,
		08193 Bellaterra (Barcelona) Spain}
	\author{Daniel Campos}
	\affiliation{Grup de F{\'i}sica Estad\'{i}stica.  Departament de F{\'i}sica.
		Facultat de Ci\`encies. Edifici Cc. Universitat Aut\`{o}noma de Barcelona,
		08193 Bellaterra (Barcelona) Spain}
	\author{Vicen\c{c} M\'{e}ndez}
	\affiliation{Grup de F{\'i}sica Estad\'{i}stica.  Departament de F{\'i}sica.
		Facultat de Ci\`encies. Edifici Cc. Universitat Aut\`{o}noma de Barcelona,
		08193 Bellaterra (Barcelona) Spain}
	\date{\today}
	
       \begin{abstract}
Random walks with stochastic resetting provides a treatable framework to study interesting features about central-place motion. In this work, we introduce non-instantaneous resetting as a two-state model being a combination of an exploring state where the walker moves randomly according to a propagator and a returning state where the walker performs a ballistic motion with constant velocity towards the origin. We study the emerging transport properties for two types of reset time probability density functions (PDFs): exponential and Pareto. In the first case, we find the stationary distribution and a general expression for the stationary mean square displacement (MSD) in terms of the propagator. We find that the stationary MSD may increase, decrease or remain constant with the returning velocity. This depends on the moments of the propagator. Regarding the Pareto resetting PDF we also study the stationary distribution and the asymptotic scaling of the MSD for diffusive motion. In this case, we see that the resetting modifies the transport regime, making the overall transport sub-diffusive and even reaching a stationary MSD., i.e., a stochastic localization. This phenomena is also observed in diffusion under instantaneous Pareto resetting. We check the main results with stochastic simulations of the process.
        \end{abstract}
        
        \maketitle
\section{Introduction}

Living organisms and moving particles in general can rarely exhibit free motion independent of environmental or internal constraints. One of these constrains consists on the presence of a privileged location which is visited with a higher frequency, either for natural or artificial reasons. For instance, in a movement ecology context, the term central-place foraging \cite{OrPe79} is often used to describe how animals seek for food near their nest. In other scenarios, such as human visual search \cite{WoHo17}, a fixed location is also used as a reference point.

From a physical point of view, this topic has been often adressed using stochastic motion models. Historically, the effect from a the central point has been modeled via an attracting potential \cite{GiAb06}. This allows the authors to analytically study the problem from a treatable perspective. However, these models do not consider the possibility of the walker returning directly to the central point, instead of (or additionally to) feeling an attraction for it. A possible mechanism that could mimic this is the random resetting of its motion to the origin.

It was not until 2011, that a simple model of stochastic motion with a strong bound to a given position was published in the physical literature \cite{EvMa11}. There, a diffusive particle is studied when it may occasionally reset its position with a constant probability and the authors find that a non-equilibrium steady state (NESS) is reached and the mean first passage time of the overall process is finite and attains a minimum in terms of the resetting rate. The existence of a NESS has been further studied for different types of motion and resetting mechanisms \cite{MoVi13,EvMa14,DuHe14,Pa15,ChSc15,CaMe15,MoVi16,MeCa16,PaKu16,NaGu16,MoMa17,Sh17,KuGu19,
MaCaMe19pp}, showing that they are not exclusive of diffusion with Markovian resets. Aside from these, other works have shown that the resetting does not always generate a NESS but transport is also possible when the resetting probability density function (PDF) is long-tailed \cite{EuMe16,MaCaMe18,BoCh19,BoCh19p} or when the resetting process is subordinated to the motion \cite{CaMe15,MeCa16}.

However, the above cited stochastic resetting models lack some realism as long as resetting is treated as an instantaneous process. Some recent papers include a quiescent period after the resetting \cite{EvMa18,MaCaMe19p}, which could mimic the time required by the walker to return to the origin. But this only serves as a partial solution to the problem since it does not consider the back-to-the-origin movement explicitly. In a slightly different context, this notion of costly resetting has also appeared in some works where search processes with restarts are studied from a Michaelis-Menten reaction scheme perspective \cite{ReUr14,RoRe15}. 

In this direction, we propose a two-state model to describe both the exploratory motion and the return to the origin, which we assume to be ballistic. We analyze the main consequences that the application of a non-instantaneous resetting has on the results known for the instantaneous case. Particularly, we derive an expression for the overall propagator to study the transport properties of the overall process for different types of resetting PDFs. We also study the statistics of the returning time in terms of the motion and the resetting PDF.

The paper is organized as follows. In Section \ref{SecII}, the model is introduced and general expressions for both the overall PDF and the overall MSD are derived. Afterwards, in Section \ref{SecIII}, the PDF of the time required by a walker to go back to the origin is studied in terms of the propagator and the resetting PDF. In Sections \ref{SecIV} and \ref{SecV} we study the particular cases of Markovian and scale-free PDFs respectively. Finally, we conclude the work in Section \ref{SecVI}.

\section{The model}
\label{SecII}

We model the dynamics of the walker by considering two different states: an \textit{exploring state} (state 1) where the motion is described by a general propagator $P(x,t)$ and a \textit{returning state} (state 2) where the motion is ballistic with velocity $v$ towards the origin defined by $x=0$. The exploring state ends at a random time according to the resetting time PDF $\varphi_R(t)$, while the returning state ends when the motion reaches the origin. At this time, the process is renewed.

Let us start by introducing the flux of particles between these two states. On one hand, assuming that the motion starts at state 1, the rate of walkers arriving at the exploring state (state 1) is

\begin{equation}
j_1(t)=\delta (t)+\int_{-\infty}^{+\infty}dz\int_0^t dt' j_2(z,t-t') \delta\left(t'-\frac{|z|}{v}\right),
\end{equation}
where the integral term is expressed in terms of the rate of the particle arriving at state 2 at point $x$ at time $t$,  $j_2(x,t)$. This term implements the probability of entering state 2 at position $z$ at any time $t-t'<t$ and subsequently reaching the origin ballistically within time $t'$ ($\delta\left(t'-|z|/v\right)$). The first term on the right hand side is the initial condition. On the other hand, the rate of walkers arriving at the returning state (state 2) reads

\begin{equation}
j_2(x,t)=\int_0^t j_1(t-t')\varphi_R(t')P(x,t')dt',
\end{equation}
which, unlike for state 1, depends explicitly on the position $x$ (i.e. the rate is not spatially uniform). Here, the probability of arriving at state 2 at position $x$ and time $t$ is the probability of having arrived at state 1 at any past time $t-t'$, times the probability that exploration finished at time $t'$, given the walker has reached position $x$. Transforming these two equations by Laplace and isolating for the rates we get the explicit expression

\begin{equation}
\hat{j}_1(s)=\left[1-\int_{-\infty}^{+\infty}dze^{-\frac{|z|}{v}s} \Pi(z,s)\right]^{-1}
\label{Eqj1}
\end{equation}
and
\begin{equation}
\hat{j}_2(x,s)=\Pi(x,s)\left[1-\int_{-\infty}^{+\infty}dze^{-\frac{|z|}{v}s} \Pi(z,s)\right]^{-1},
\label{Eqj2}
\end{equation}
where we have introduced the notation
\begin{equation}
\Pi(x,s)\equiv \mathcal{L}[\varphi_R(t)P(x,t)],
\label{Pi}
\end{equation}
and
\[
\hat{f}(s)=\mathcal{L}[f(t)]=\int_0^\infty dt e^{-st}f(t)dt
\]
is the Laplace transform of $f(t)$.

Let us now introduce the spatial dynamics for each of the states. We write the global propagator as $\rho(x,t)=\rho_1(x,t)+\rho_2(x,t)$, i.e. the overall motion is described by the combination of the propagators in states 1 and 2. Defining the probability that no reset has occurred until time $t$ as $\varphi_R^*(t)=\int_t^\infty \varphi_R(t')dt'$, the probability of finding the walker at the exploring state at point $x$ and time $t$ can be written as

\begin{equation}
\rho_1(x,t)=\int_0^tj_1(t-t')\varphi_R^*(t')P(x,t')dt'.
\end{equation}
Transforming by Laplace and inserting Eq.\eqref{Eqj1}, we have 

\begin{equation}
\hat{\rho}_1(x,s)=\Pi^*(x,s)\left[1-\int_{-\infty}^{+\infty}dze^{-\frac{|z|}{v}s} \Pi(z,s)\right]^{-1},
\end{equation}
with

\begin{equation}
\Pi^*(x,s)\equiv \mathcal{L}[\varphi_R^*(t)P(x,t)].
\label{Pia}
\end{equation}

For the sake of simplicity we will restrict to symmetric random walks, i.e. $P(x,t)=P(-x,t)$. Then, for the returning state  (we consider $x>0$, which is labeled by a plus sign in the following, to later generalize this result for $x<0$) we have that

\begin{equation}
\rho_2^+(x,t)=\int_{x}^{+\infty}dz\int_0^t dt' j_2(z,t-t')\delta(x-z+vt').
\label{EqProp2_0}
\end{equation}
This is, to be at position $x>0$ at time $t$ in the returning state, the motion had to reach this state at a previous time $t-t'$ and position $z>x$ and, during the time $t'$, the returning ballistic motion must have gone from $z$ to $x$. Transforming by Laplace, we get

\begin{equation}
\hat{\rho}_2^+(x,s)=\frac{e^{\frac{x}{v}s}}{v}\int_{x}^{+\infty}dz e^{-\frac{z}{v}s}\hat{j}_2(z,s).
\label{EqProp2_1}
\end{equation}
Now, introducing Eq.\eqref{Eqj2} and generalizing it for the negative positions (i.e. $x\rightarrow |x|$ in Eq.\eqref{EqProp2_1}), we get the general expression for the propagator in the state $i=2$

\begin{equation}
\hat{\rho}_2(x,s)=\frac{1}{v}e^{\frac{|x|}{v}s}\frac{\int_{|x|}^{+\infty}dze^{-\frac{|z|}{v}s} \Pi(z,s)}{1-\int_{-\infty}^{+\infty}dze^{-\frac{|z|}{v}s} \Pi(z,s)}.
\end{equation}

Finally, putting both propagators together, the overall propagator $\hat{\rho}(x,s)=\hat{\rho}_1(x,s)+\hat{\rho}_2(x,s)$ turns out to be
\begin{equation}
\hat{\rho}(x,s)=\frac{\Pi^*(x,s)+\frac{1}{v}e^{\frac{|x|}{v}s}\int_{|x|}^{+\infty}dze^{-\frac{|z|}{v}s} \Pi(z,s)}{1-\varphi(s)},
\label{EqPropRealL}
\end{equation}
where 
\begin{eqnarray}
\varphi(s)&=&\int_{-\infty}^{+\infty}dze^{-\frac{|z|}{v}s} \Pi(z,s)\nonumber\\ 
&=&\frac{vs}{\pi}\int_{-\infty}^{+\infty}dk \frac{\Pi(k,s)}{k^2v^2+s^2},
\label{EqDefPhi}
\end{eqnarray}
with $\Pi(k,s)$ the Fourier-Laplace transform of $\Pi(x,t)$. 

Therefore, for a general resetting PDF $\varphi_R(t)$ and propagation $P(x,t)$, one can in principle find the PDF of the overall motion process. In the numerator we have the contributions of the exploring and returning states clearly separated. However, the first term, corresponding to the exploring state, is velocity-dependent since in the denominator $v$ appears explicitly. Therefore, both the exploring and returning states are actively modified by the finiteness of the resetting velocity.
Transforming Eq. (\ref{EqPropRealL}) by Fourier one finally gets, after some calculations,

\begin{eqnarray}
\rho (k,s)&=&\frac{1}{1-\varphi(s)}\left[\Pi ^*(k,s)+s\frac{\Pi(k,s)-\varphi(s)}{k^2v^2+s^2}\right.\nonumber\\ 
&+&\left.\frac{2kv}{k^2v^2+s^2}\int_0^\infty dz\sin(kz) \Pi (z,s)\right],
\label{rofl}
\end{eqnarray}
where

\begin{equation}
\Pi(k,s)=\int_0^\infty e^{-st}\varphi_R(t)P(k,t)dt.
\label{Piks}
\end{equation}
One can check that $\rho (k=0,s)=1/s$, so $\rho (x,t)$ is conveniently normalized.

\subsection{MSD}
The MSD can be obtained as

\begin{equation}
\langle \hat{x}^2(s)\rangle =-\left[\frac{\partial ^2\rho(k,s)}{\partial k^2}\right]_{k=0}.
\label{msd0}
\end{equation}
Using Eq. (\ref{rofl}), after some calculations one obtains

\begin{equation}
\langle \hat{x}^2(s)\rangle =\frac{\mathcal{L}\left[ \varphi_R^*(t)\langle x^2(t)\rangle_P \right]+ \frac{1}{s} \mathcal{L}\left[ \varphi_R(t)\langle x^2(t)\rangle_P \right]- 2\frac{v}{s^3} \Phi(s)}{1-\varphi(s)},
\label{EqMSDLap}
\end{equation}
where
\begin{equation}
\Phi(s)\equiv s\mathcal{L}\left[ \varphi_R(t)\langle |x|(t)\rangle_P \right]+v[ \varphi(s)-\varphi_R(s)]
\end{equation}
and

\begin{eqnarray}
\langle x^2(t)\rangle_P&=&\int_{-\infty}^{\infty}x^2P(x,t) dx,  \nonumber\\ 
\langle |x|^n(t)\rangle_P&=&2\int_0^\infty x^nP(x,t) dx,
\label{def}
\end{eqnarray}
with $n=1,2,...$. In the large time limit, that is, $s\rightarrow 0$ we can simplify Eq. (\ref{EqMSDLap}) keeping the leading order terms. One gets, after some algebra

\begin{equation}
\langle \hat{x}^2(s)\rangle \simeq \frac{\mathcal{L}\left[ \varphi_R^*(t)\langle x^2(t)\rangle_P \right]+\frac{1}{3v}\mathcal{L}\left[ \varphi_R(t)\langle |x|^3(t)\rangle_P \right]}
{1-\varphi_R(s)+\frac{s}{v}\mathcal{L}\left[ \varphi_R(t)\langle |x|(t)\rangle_P \right]}.
\label{msdap}
\end{equation}
Therefore, a finite returning velocity makes the overall asymptotic MSD depend explicitly on the first and third absolute-value moments of the propagator, in addition to the second moment. Crucially, this dependence disappears when the limit $v\rightarrow \infty$ is taken to recover the equivalent expression for instantaneous resetting \cite{MaCaMe18}.

In Sections \ref{SecIV} and \ref{SecV}, we explore this and more properties on particular scenarios to analyze in more detail the effect that a finite resetting velocity has at long times, comparing our model to some of the most significant results obtained for instantaneous resets.

\subsection{Returning time PDF}
\label{SecIII}
After a resetting event, the random walker is forced to interrupt its motion and go back to the origin with constant velocity $v$. During the time required to go back to the origin and restart its motion, the walker remains in a state which hinders the overall propagation. Similar models have been studied in the literature, where the walker is forced to remain inactive at the origin after resetting \cite{EvMa18,MaCaMe19p}. Nevertheless, in none of these models the duration of the non-propagating period is considered to depend explicitly on the motion of the walker, which is a natural correlation to be taken into account.

In \cite{MaCaMe18}, the asymptotic behavior of the overall MSD is studied for a stochastic motion process which resets its position at times distributed according to a resetting PDF, after which remains at the origin during a period determined by a residence time PDF. There, the scaling of the overall MSD is shown to strongly depend on the resetting and the residence time PDFs. In the current model, while the resetting PDF is explicitly introduced, the PDF of the non-propagating period emerges from the combination of both the resetting time PDF $\varphi_R(t)$ and the stochastic motion performed by the walker, i.e. the propagator $P(x,t)$. Therefore, it is convenient to study the stochastic properties of the returning time for a later analysis of the overall transport properties of the system.

In general, the PDF of the returning time can be written as

\begin{equation}
\varphi_{r}(t)=\int_{-\infty}^{+\infty}dx\ \delta\left(t-\frac{|x|}{v}\right)\int_0^\infty dt' P(x,t')\varphi_R(t'),
\end{equation}
which is the probability of the exploratory motion ending at a given position $x$ ($t'$ integral) times the probability of the returning state to last at time $t$ from this position (delta term), averaged over all the possible positions. The integral in $x$ can be performed to give a general expression for the returning time PDF in terms of both the propagator and the resetting PDF:

\begin{equation}
\varphi_{r}(t)=2v\int_0^\infty P(vt,t')\varphi_R(t')dt'=2v\ \Pi(vt,0),
\label{EqRetPDFgeneral}
\end{equation}
with $\Pi(x,s)$ defined from Eq.\eqref{Pi}. An equation for the n-th moment of the returning time PDF can be also found by multiplying both sides of Eq.\eqref{EqRetPDFgeneral} by $t^n$ and integrating over $t>0$:

\begin{equation}
\langle t_{r}^n\rangle =\frac{1}{v^n}\int_0^{\infty} \varphi_R(t)\langle |x(t)|^n\rangle_P dt.
\label{EqRetMTgeneral}
\end{equation}

It is interesting to observe that the contribution of the motion to the $n$-th moment of the returning time PDF comes exclusively from the $n$-th absolute moment of the motion propagator. In the following we dig into these results for two different types of resetting time PDFs.

\subsubsection{Markovian resetting}

\begin{figure*}
\includegraphics[scale=0.55]{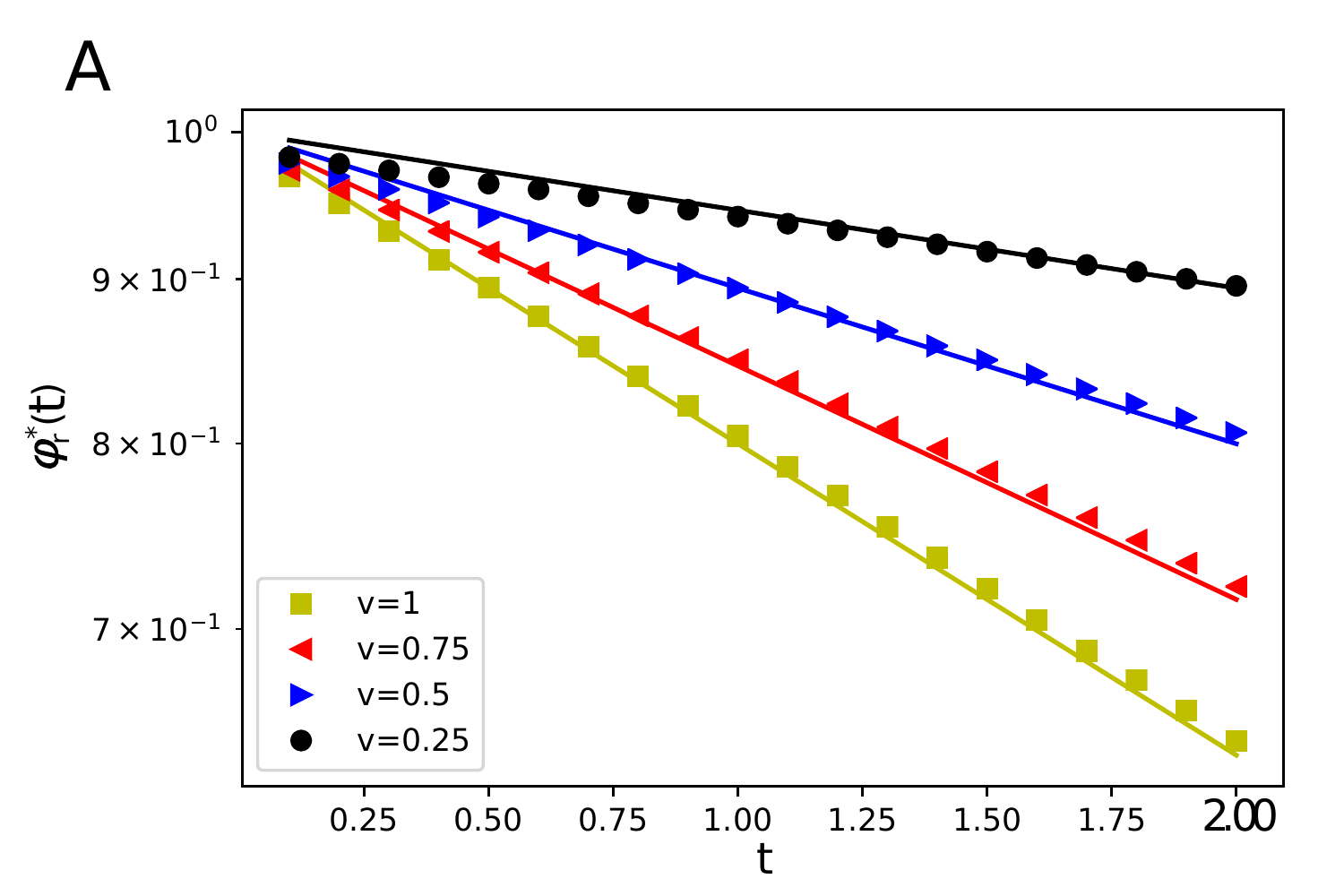}
\includegraphics[scale=0.55]{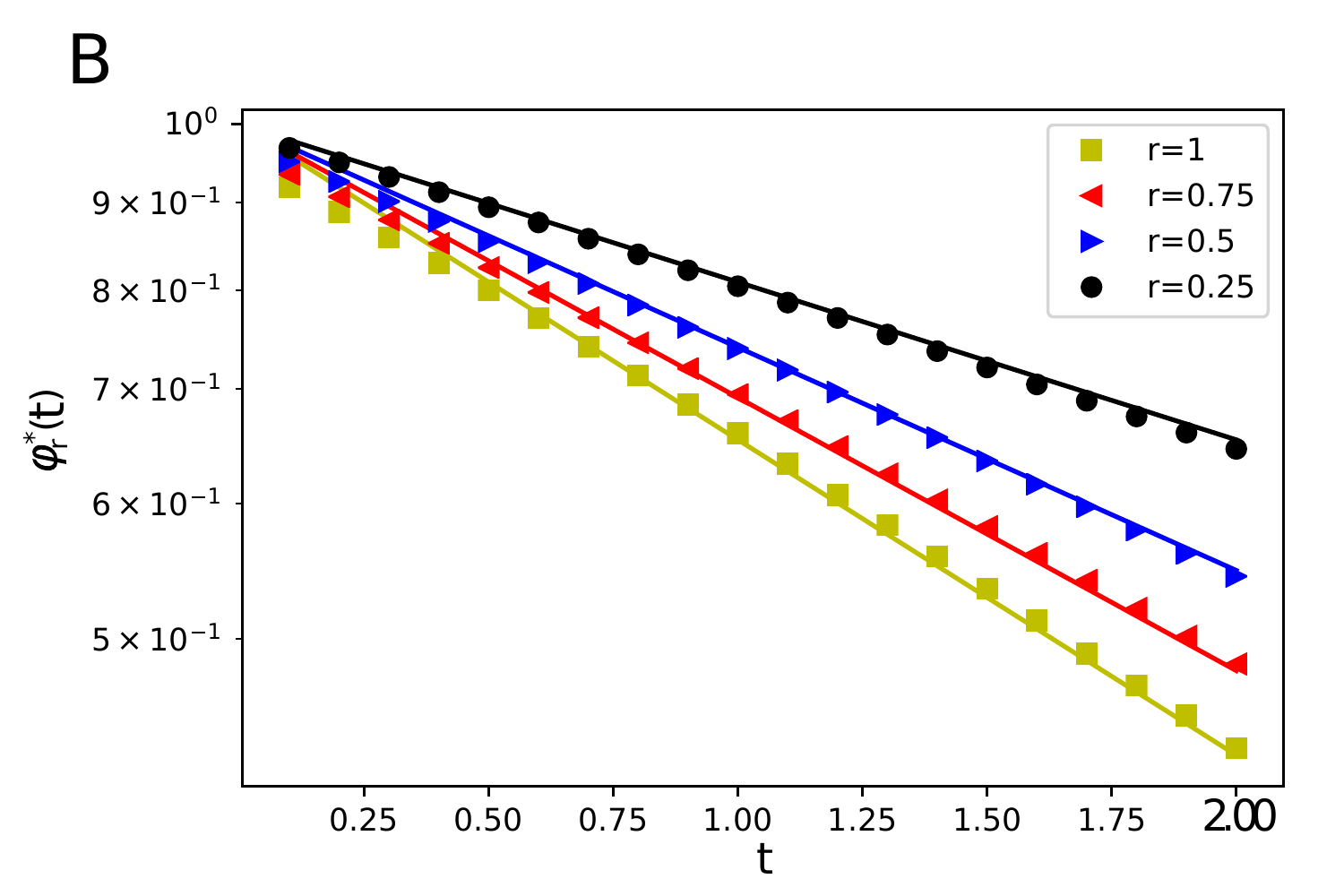}
\caption{On the left (panel A), the returning time survival probability $\varphi_{r}^*(t)=\int_t^\infty \varphi_{r}(t')dt'=e^{-v\sqrt{\frac{r}{D}}t}$ for diffusion with $D=5$ under exponential resetting with $r=0.25$ is shown for different values of the returning velocity $v$ in a stochastic simulation of the process. On the right (panel B), the returning time distribution for the same motion is shown with exponential resetting for different $r$ and returning velocity $v=1$. In both cases, the PDF is represented in a log-linear axis and the exponential behavior can be clearly observed to be in agreement with the corresponding analytical prediction (solid lines).}
\label{FigExpReturning}
\end{figure*}

As a first case, we analyze the returning time PDF when the resetting is equally probable at any time. This consists on choosing a reset time PDF of the form
\begin{equation}
\varphi_R(t)=r e^{-rt}.
\label{ExpDist}
\end{equation} 
The general expressions derived above reduce to 

\begin{equation}
\varphi_{r}(t)=2vr\ \hat{P}(vt,r)
\label{EqRetPDFexp}
\end{equation}
and
\begin{equation}
\langle t_{r}^n\rangle =\frac{r}{v^n}\langle |\hat{x}(r)|^n\rangle_P.
\label{EqRetMTexp}
\end{equation}
 Notably, if the n-th moment of the motion has a finite Laplace transform, the $n$-th moment of the returning time PDF converges to a finite value. This includes, for instance, all the cases where the moments grow as a power law in time. Between these, let us consider diffusive motion with a Gaussian propagator of the form $P(x,t)=\exp(-x^2/4Dt)/\sqrt{4\pi Dt}$. After some simple calculations, the returning time PDF can be found to be exponential

\begin{equation}
\varphi_{r}(t)=v\sqrt{\frac{r}{D}}e^{-v\sqrt{\frac{r}{D}}t}.
\label{EqRetEG}
\end{equation}
This result has been compared to stochastic simulations of the process and it has been seen to be in agreement with them (Figure \ref{FigExpReturning}).

Therefore, for diffusive Gaussian motion, when the resetting PDF is exponential, the duration of the returning state also follows an exponential distribution. In other words, the transition from the returning state to the exploring state is also markov process and happens at a constant rate which can be identified to be 

\[
r_{r}=v \sqrt{\frac{r}{D}}.
\]
Remarkably, the return-to-explore transition rate depends on the explore-to-return rate as a square root. This is, by increasing $r$, the rate of the return-to-explore transition will be less increased than the rate of the explore-to-return transition. It is also interesting to observe that the returning rate depends linearly with the velocity $v$. So, aside from the weight factor $\sqrt{r/D}$ related to the propagation ability of the motion, the velocity can be interpreted as the actual transition rate from the returning to the exploring state.

\subsubsection{Scale-free resetting}

Let us now explore the case where the resetting PDF is scale free, meaning that in the long time regime it decays as

\begin{equation}
\varphi_R(t)\sim t^{1+\gamma},
\label{EqDecayPareto}
\end{equation}
with $\gamma$ a real, positive number. To do so, we employ a particular form of the resetting PDF, being a Pareto distribution of the form 

\begin{equation}
\varphi_R (t)=\frac{\gamma r}{\left(1+r t\right)^{1+\gamma}},
\label{pareto}
\end{equation}
with $\gamma >0$. As a difference with the exponential PDF, this choice allows us to study a resetting PDF with diverging moments, having that the m-th moment will exist whenever $\gamma >m$. In particular, here we will consider diffusive motion. Introducing the resetting PDF and the propagator to Eq.\eqref{EqRetPDFgeneral}, one can formally write
\begin{equation}
\varphi_{r}(t)=v\gamma\sqrt{\frac{ r}{ \pi D}}\Gamma\left(\gamma+\frac{1}{2}\right)U\left(\gamma+\frac{1}{2},\frac{1}{2};\frac{r v^2}{4D}t^2\right),
\label{EqRetPDFPareto}
\end{equation}
where $U(a,b;z)$ is the Tricomi confluent hypergeometric function or confluent hypergeometric function of the second kind (see Section 13 in \cite{Ab74}). Now, from the long time behavior of the Tricomi function we can obtain the decaying of the returning PDF:
\begin{equation}
\varphi_{r}(t)\sim t^{-(1+2\gamma)}\;\;\text{as} \; t\rightarrow\infty.
\end{equation}
Therefore, the returning PDF always decays faster than the resetting PDF (see Eq.\eqref{EqDecayPareto}).

Let us now derive the moments of the returning PDF. To do so, we employ Eq.\eqref{EqRetMTgeneral} and use the expression for the $n$-th absolute moment of a Gaussian distribution
\begin{equation}
\langle |x(t)|^n\rangle =\Gamma\left( \frac{n+1}{2}\right)\sqrt{\frac{(4D)^n}{\pi}}t^{\frac{n}{2}}.
\end{equation}
This way, one can express the moments of the returning PDF as

\begin{equation}
\langle t_{r}^n\rangle =2v \gamma r U\left(\frac{n}{2}+1,\frac{n}{2}+1-\gamma;0\right).
\end{equation}
Now, the Tricomi confluent hypergeometric function $U(a,b;z)$ at $z=0$ does not converge when $b<1$. This establishes a condition for the moments of the returning time PDF to exist, being that the $n$-th moment of $\varphi_{r}(t)$ exist when 

\begin{equation}
\gamma>\frac{n}{2}.
\label{EqCondConvergence}
\end{equation}
In particular, the returning time will have a finite mean ($n=1$) only if $\gamma>1/2$. Therefore, for $1/2<\gamma <1$, despite the exploration (resetting) time has diverging mean value, the mean returning time to the origin will be finite.

\begin{figure}
\includegraphics[scale=0.55]{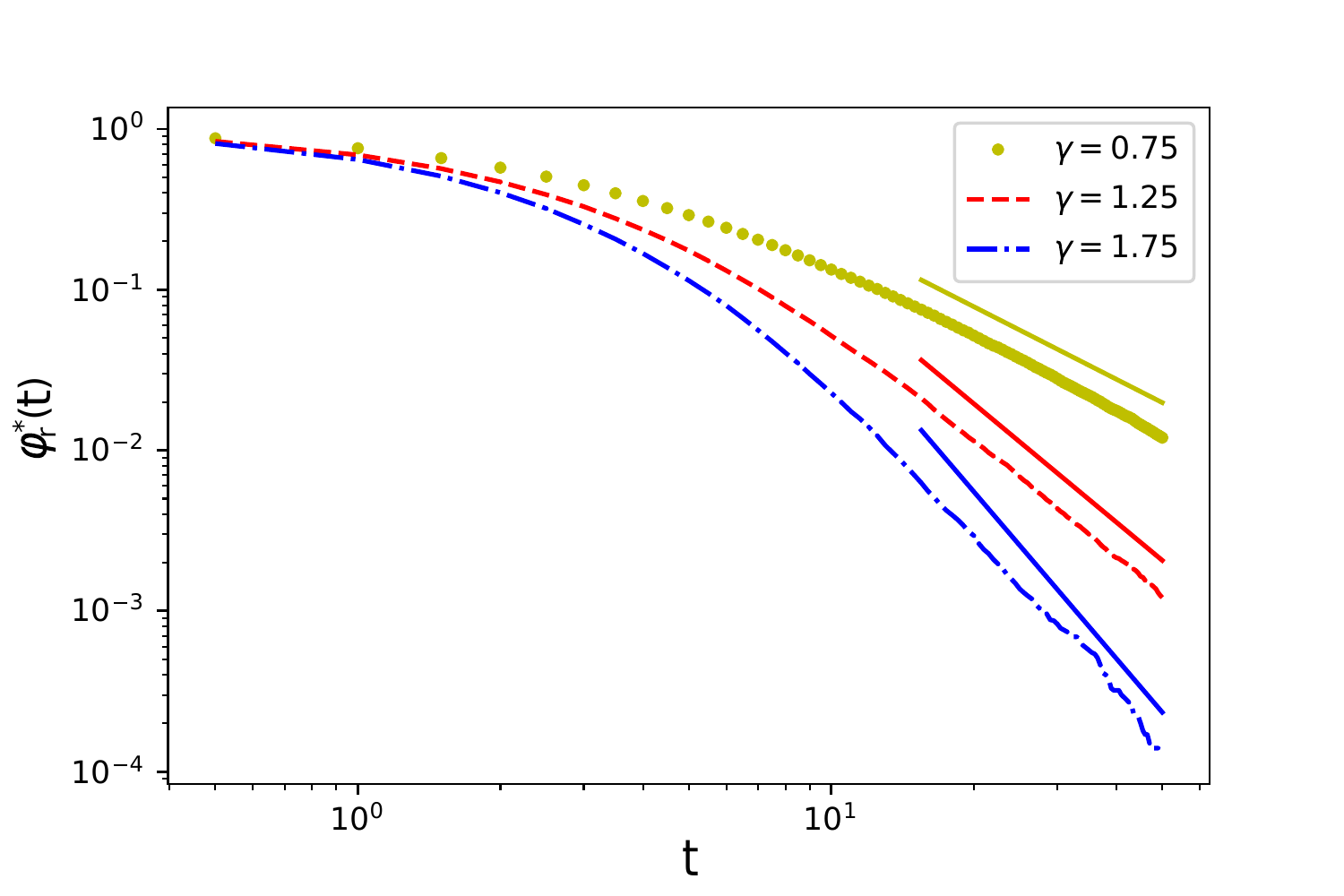}
\caption{The returning time survival probability $\varphi_{r}^*(t)=\int_t^\infty \varphi_{r}(t')dt'\sim 1/t^{2\gamma}$ is shown for three different simulations with the corresponding $\gamma$ parameters. A diffusion process with $D=5$ with resets with $r=1$ and returning velocity $v=1$ have been employed. Straight solid lines of slope $-2\gamma$ have been included as a guide for the eye.}
\label{FigParReturning}
\end{figure}

\section{Markovian resetting}
\label{SecIV}
In this Section we analyze the large time behavior of the propagator $\rho (x,t)$ and the MSD $\langle x^2(t)\rangle$ for a random walker under non-instantaneous resettings when the resetting times are drawn from an exponential PDF (Eq.\eqref{ExpDist}). This has been the most studied case in the literature and various random walks have been tested under this type of resets. For this particular case, one has from Eqs. \eqref{Pi} and \eqref{Pia} that
\begin{equation}
\Pi(x,s)=r \hat{P}(x,s+r)
\end{equation}
and
\begin{equation}
\Pi^*(x,s)=\hat{P}(x,s+r).
\end{equation}
If $P(x,t)$ has non-diverging moments, in the large time limit (small $s$), Eq.\eqref{EqDefPhi} reads

\begin{equation}
\varphi(s) = 1-sr-\frac{sr}{v}\langle |\hat{x}(s=r)|\rangle_P +O(s^2)
\end{equation}
and the overall propagator reaches the NESS

\begin{equation}
\rho_{s}(x)= \frac{\hat{P}(x,s=r)+\frac{r}{v}\int_{|x|}^\infty \hat{P}(z,s=r)dz}{\frac{1}{r}+\frac{r}{v}\langle |\hat{x}(s=r)|\rangle _P}.
\label{EqNESSexpo}
\end{equation}
Here we have made use of Eq. (\ref{rofl}) and inverted by Fourier. It is interesting to note that this is a general result for any symmetric propagator $P(x,t)$ with finite moments under exponential resetting times. 

Despite of the tedious calculations employed to reach it, this result is extremely simple to interpret from a physical point of view. We have two different contributions to the PDF at the NESS: the first term in the numerator accounts for the propagation of the stochastic motion, while the second term accounts for the walkers returning to the origin at a finite velocity, after suffering a reset at a position $|z|>|x|$. In fact, taking the limit $v\rightarrow \infty$, corresponding to instantaneous resets, one recovers the result recently found in \cite{MaCaMe18} for Markovian resetting applied to a general propagator $P(x,t)$.

In \cite{MaCaMe19p}, a model with instantaneous resetting followed by a random residence period at the origin was studied. There, it was seen that considering a residence period does not modify the shape of the NESS. In particular, this occurs if one considers the returning time distribution obtained in Eq.\eqref{EqRetPDFexp}. Contrarily, in Eq.\eqref{EqNESSexpo} one can see that considering a finite returning velocity does modify the shape of the NESS. Therefore, from the model in \cite{MaCaMe19p} with the distribution in Eq.\eqref{EqRetPDFexp} one cannot emmulate the NESS arising from the non-instantaneous resetting model.

More specifically, for a diffusive random walks the propagator follows a Gaussian distribution, which in the Laplace space takes the form $\hat{P}(x,s)=\sqrt{1/4sD}\exp (-|x|\sqrt{s/D})$. Inserting this result into Eq. (\ref{EqNESSexpo}) we obtain 

\begin{equation}
\rho_{s}(x)=\sqrt{ \frac{r}{4D}}e^{-|x|\sqrt{r/D}}
\label{EqNESSexpo}
\end{equation}
which is independent of the returning velocity. This is a well-known result already obtained for diffusing particles under markovian instantaneous resettings  \cite{EvMa11}.
Regarding the MSD, inserting the exponential distribution for the resetting times into Eq. (\ref{msdap}) one readily finds that the MSD tends, in the $t\rightarrow \infty$ limit, to
 
\begin{equation}
\langle x^2\rangle_{s}= \frac{\langle \hat{x}(s=r)^2\rangle_P+\frac{r}{3v}\langle |\hat{x}(s=r)|^3\rangle_P}{\frac{1}{r}+\frac{r}{v}\langle |\hat{x}(s=r)|\rangle _P}.
\label{EqMSDexpo}
\end{equation}
Notably, the overall stationary MSD does not depend only on the MSD of the motion $\langle x^2\rangle_P$ as does when the resetting is instantaneous \cite{EvMa11,MaCaMe18} but also on its first and third absolute-value moments. 

Let us analyze how the overall MSD depend on the velocity. An extreme (either maximum or minimum) for the MSD in terms of the velocity $v$ does not exist. Nevertheless, depending on the relative values of the three first moments of the motion, the overall MSD may increase, decrease or remain constant with $v$. General conditions can be indeed established for any propagator under exponential resetting PDF. Thus, we have that the MSD decreases with $v$ if

\begin{equation}
\langle |\hat{x}(s=r)|^{3}\rangle_P>3r\langle |\hat{x}(s=r)|\rangle_P \langle \hat{x}(s=r)^{2}\rangle_P,
\label{c1}
\end{equation}
and the MSD remains constant with $v$. It is independent on the velocity if

\begin{equation}
\langle |\hat{x}(s=r)|^{3}\rangle_P=3r\langle |\hat{x}(s=r)|\rangle_P \langle \hat{x}(s=r)^{2}\rangle_P.
\label{c2}
\end{equation}
In this case, it reduces to the instantaneous resetting value $\langle x^2\rangle_{s}= r \langle \hat{x}(s=r)^2\rangle_P $ \cite{MaCaMe18}.
Finally, the MSD increases with $v$ if

\begin{equation}
\langle |\hat{x}(s=r)|^{3}\rangle_P<3r\langle |\hat{x}(s=r)|\rangle_P \langle \hat{x}(s=r)^{2}\rangle_P.
\label{c3}
\end{equation}
Since $\langle x^2\rangle_{s}$ is directly related to the width of the NESS  distribution, its shape is affected by the returning velocity $v$ in those cases where $\langle x^2\rangle_{s}$ depends on $v$, that is, when one of the conditions (\ref{c1}) or (\ref{c3}) is fulfilled. 
Let us be more specific and consider two cases: the propagators for Brownian motion and for fractional Brownian motion (fBm). In the former case $P(x,t)=[4\pi Dt]^{-1/2}\exp(-x^2/4Dt)$ and the moments involved in Eq. (\ref{EqMSDexpo}) are, in the Laplace space, given by

\begin{eqnarray}
\langle \hat{x}(s)^{2}\rangle_P &=& 2D/s^2\nonumber\\ 
\langle |\hat{x}(s)|\rangle_P&=&\sqrt{D}/s^{3/2}\nonumber\\ 
\langle |\hat{x}(s)|^{3}\rangle_P &=&6D^{3/2}/s^{5/2}.
\label{moments}
\end{eqnarray}
Introducing these results into Eq. (\ref{EqMSDexpo}) and simplifying one finds $\langle x^2\rangle_{s}=2D/r$, so that the overall MSD in the large time limit does not depends on $v$. Note that in this case, condition (\ref{c2}) is fulfilled and the NESS distribution does not depend on $v$. 

For a fBm, the propagator is given by $P(x,t)=[4\pi Kt^\alpha]^{-1/2}\exp(-x^2/4Kt^\alpha)$ and the moments we are interested in are
\begin{eqnarray}
\langle \hat{x}(s)^{2}\rangle_P &=& 2K\Gamma(1+\alpha)/s^{1+\alpha}\nonumber\\ 
\langle |\hat{x}(s)|\rangle_P&=&\sqrt{4K}\Gamma(1+\frac{\alpha}{2})/\sqrt{\pi} s^{1+\frac{\alpha}{2}}\nonumber\\ 
\langle |\hat{x}(s)|^{3}\rangle_P &=&(4K)^{\frac{3}{2}}\Gamma(1+\frac{3\alpha}{2})/\sqrt{\pi} s^{1+\frac{3\alpha}{2}}.
\label{moments2}
\end{eqnarray}
in the Laplace space. It is easy to check from (\ref{c1})-(\ref{c3})  that, unlike the normal diffusion, the overall MSD increases with $v$ when $0<\alpha <1$ while it decreases with $v$ when $1<\alpha<2$. When $\alpha =1$ we recover the normal diffusion case and the MSD does not depend on $v$.  
The overall MSD is found from Eq. (\ref{EqMSDexpo}) and Eq. (\ref{moments2}), which lead to

\begin{equation}
\langle x^2\rangle_{s}=\frac{2K\Gamma (1+\alpha)}{r^{\alpha}}\frac{1+ \frac{2\Gamma(1+\frac{3\alpha}{2})}{3\sqrt{\pi}\Gamma(1+\alpha)}  \frac{\sqrt{4K}r^{1-\frac{\alpha}{2}}}{v}  }{1+   \frac{\Gamma(1+\frac{\alpha}{2})}{\sqrt{\pi}}  \frac{\sqrt{4K}r^{1-\frac{\alpha}{2}}}{v} }
\label{msdfbm}
\end{equation}

\begin{figure}
\includegraphics[scale=0.625]{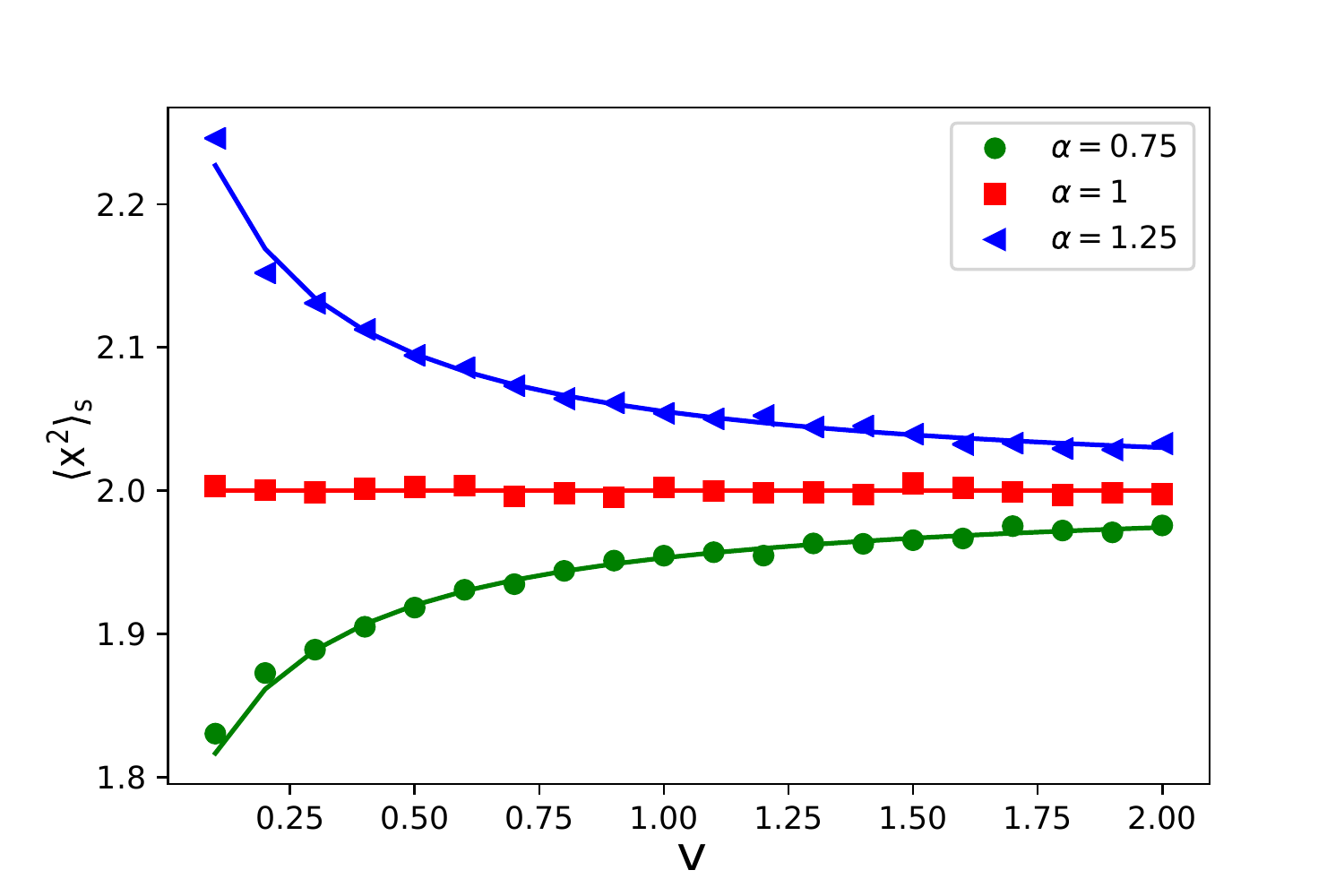}
\caption{The results of a stochastic simulation of the stationary MSD for two fractional Brownian motion with $\alpha=0.75$ and $\alpha=1.25$, and a normal Brownian motion with $\alpha=1$ with non-instantaneous resetting ($r=0.2$) are presented in terms of the returning velocity $v$. The multiplicative constant has been chosen to be $K=r^\alpha/\Gamma(1+\alpha)$ in order to have the same instantaneous resetting limit (large $v$) for all of them. The solid lines are the corresponding analytical predictions from Eq.\eqref{msdfbm}.}
\label{FigMSDvsV}
\end{figure}

In Figure \ref{FigMSDvsV} we show three representative cases of this result compared to stochastic simulations. There, one can see that for motion processes which are prone to stay near the origin, as fBm with $\alpha <1$, the stationary MSD increases with $v$. Otherwise, for processes which quickly move away from the origin, as fBm with $\alpha >1$, a larger returning velocity $v$ makes the stationary MSD decrease. Finally, when $\alpha =1$, which is the case of Gaussian propagator, the overall stationary MSD is shown to not depend on the returning velocity.

\section{Scale-free resetting}
\label{SecV}
In this section we consider a Gaussian propagator and a Pareto for the resetting time PDF. For practical examples of phenomena that may generate scale-free reset times we refer the reader to \cite{NaGu16}. The interest of this distribution is that the exponent $\gamma$ controls the existence of finite moments. When $0<\gamma \leq 1$ the distribution lacks of moments and behaves like a Mittag-Leffler distribution in the large time limit, i.e., it decays as $t^{-1-\gamma}$. However, when $1<\gamma\leq 2$ only the first moment is finite and there exists a characteristic resetting rate, when $2 <\gamma\leq 3$ only the first and second moments exist and so on and so forth.

\begin{figure*}
\includegraphics[scale=0.57]{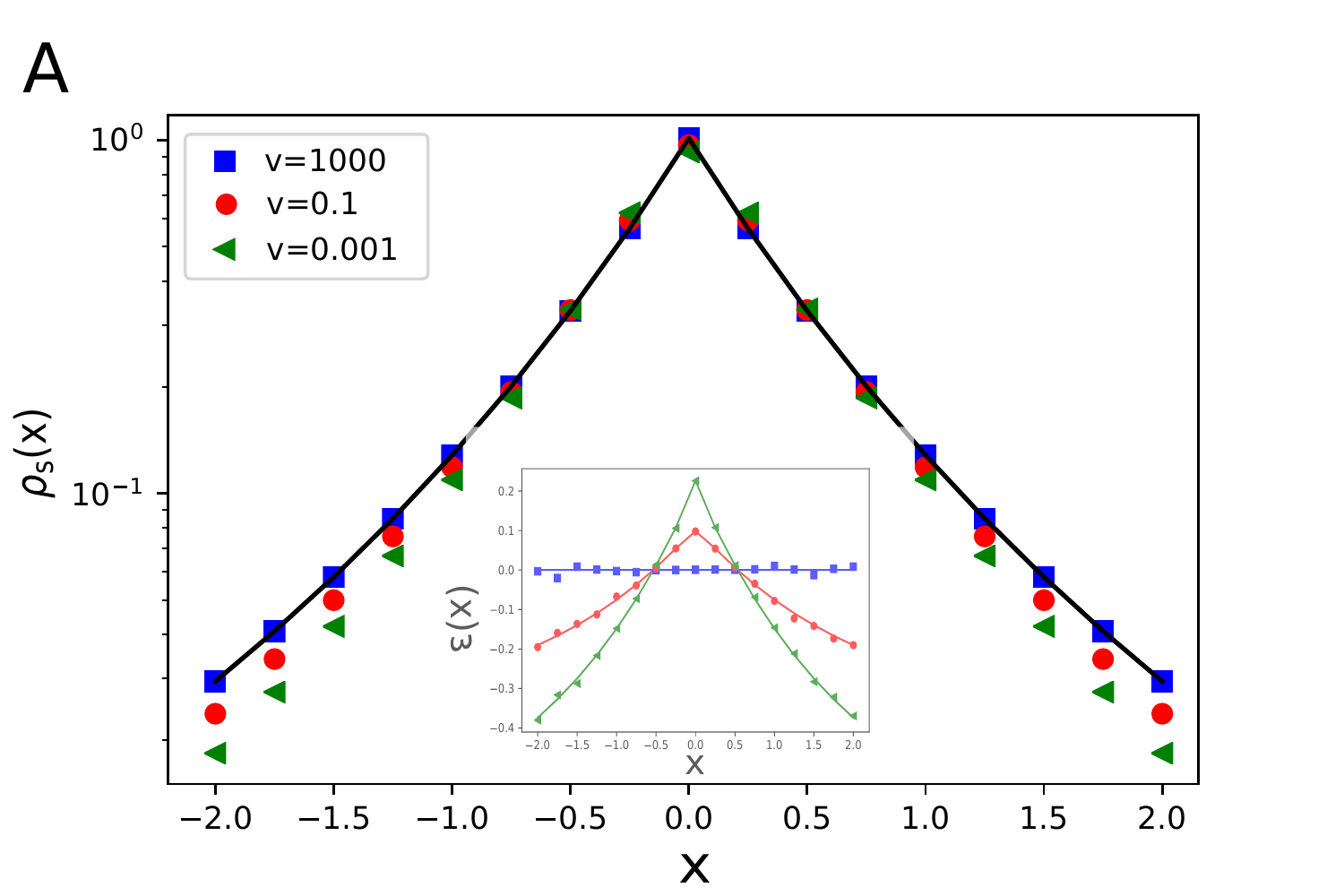}
\includegraphics[scale=0.57]{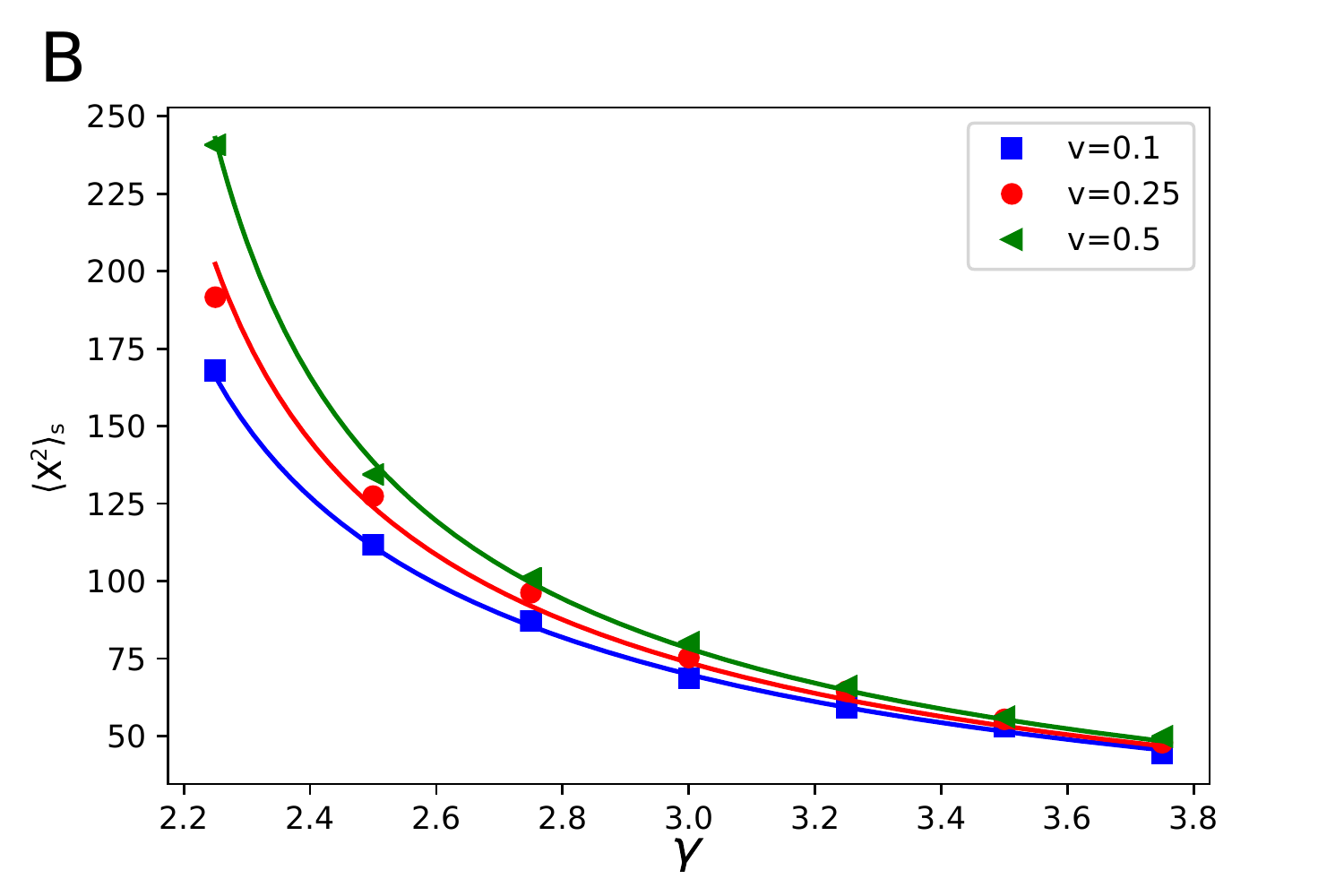}
\caption{Diffusive motion resetted according to a Pareto PDF with $r=0.2$ has been stochastically simulated. On the left (panel A), we show the NESS from simulations with exponent $\gamma=2.25$ in the Pareto PDF, diffusion constant $D=0.05$ and different returning velocities. The solid black line corresponds to the instantaneous resetting case, which can be obtained by taking the $v\rightarrow \infty$ limit in Eq.\eqref{rspareto}. As an inset, we include the relative distance between the different non-instantaneous resetting NESS and the instantaneous resetting NESS for simulations (squares, dots and triangles) and the corresponding analytical distribution from Eq.\eqref{rspareto} (solid curves). On the right (panel B), the stationary MSD is ploted in terms of $\gamma$ for three different values of the returning velocity and with diffusion constant $D=5$.}
\label{Fig4}
\end{figure*}

Our first goal is to study the large time behavior of the overall distribution. To this end we take the limit $s\rightarrow 0$ in  Eq. (\ref{EqPropRealL}). For a diffusive propagator for the exploring state and the resetting times PDF in Eq. (\ref{pareto}) we obtain that the NESS is reached when $\gamma >1$ and it reads 

\begin{eqnarray}
\rho_s(x)&=&
\frac{(\gamma-1)/\sqrt{\frac{4\pi D}{r}}}{1+\frac{\sqrt{D r}}{v}\frac{\Gamma\left(\gamma-\frac{1}{2}\right)}{\Gamma(\gamma-1)}}
\left[ \Gamma\left(\gamma-\frac{1}{2}\right)U\left(\gamma-\frac{1}{2},\frac{1}{2};\frac{rx^2}{4D} \right)\right. \nonumber\\ 
&+&\left. \frac{r |x|}{2v}\Gamma(\gamma+\frac{1}{2})U\left(\gamma+\frac{1}{2},\frac{3}{2};\frac{rx^2}{4D}  \right)\right] .
\label{rspareto}
\end{eqnarray}
This has been shown to be in agreement with the results from stochastic simulations of the process (see inset in Figure \ref{Fig4}A). As for the stationary state with Markovian resetting, this result shows that the returning state modifies the shape of the NESS. Particularly, in Figure \ref{Fig4}A one can see that the NESS becomes wider when increasing the velocity, showing an asymptotic tendency to the instantaneous resetting NESS. 

Our second goal is to find the overall MSD by using Eq. (\ref{msdap}). To this end we need to compute separately the Laplace transforms that appear in this equation. The Laplace transform of the Pareto PDF is

\begin{equation}
\hat{\varphi}_R (s)=\gamma U(1,1-\gamma;s/r)
\label{paretos}
\end{equation}
Analogously,

\begin{eqnarray}
\mathcal{L}\left[ \varphi_R(t)\langle |x(t)|\rangle_P \right]&=&\gamma \sqrt{\frac{D}{r}}U\left( \frac{3}{2},\frac{3}{2}-\gamma; s\frac{s}{r}\right) \nonumber\\ 
\mathcal{L}\left[ \varphi_R(t)\langle |x(t)|^3\rangle_P \right]&=&6\gamma \left( \frac{D}{r}\right) ^{\frac{3}{2}}U\left( \frac{5}{2},\frac{5}{2}-\gamma; \frac{s}{r}\right)
\label{lt13}
\end{eqnarray}    
where we made use of Eq. (\ref{moments}). Finally, since the survival PDF is $\varphi _R^*(t)=(1+rt)^\gamma$ we get

\begin{equation}
\mathcal{L}\left[ \varphi_R^*(t)\langle x(t)^2\rangle_P \right]=2D/r^2U\left(2,3-\gamma; s/r\right).
\label{lt2}
\end{equation}
Taking into account the asymptotic expansions for small arguments of the Tricomi functions (see Section 13 in \cite{Ab74} for details) we find that, for $s/r \ll 1$,
\begin{equation}
1-\hat{\varphi} _R(s)\sim \left\{
\begin{matrix} s^{\gamma},\ \ \gamma \leq 1\\
{}\\
s,\ \ \gamma > 1\\
\end{matrix} \right. 
\label{fap}
\end{equation}
and from (\ref{lt13}) we obtain

\begin{equation}
\mathcal{L}\left[ \varphi_R(t)\langle |x(t)|\rangle_P \right]\sim \left\{
\begin{matrix} s^{\gamma-\frac{1}{2}},\ \ \gamma \leq \frac{1}{2}\\
{}\\
s^0,\ \ \gamma > \frac{1}{2},\\
\end{matrix} \right. 
\label{m1ap}
\end{equation}

\begin{equation}
\mathcal{L}\left[ \varphi_R^*(t)\langle x(t)^2\rangle_P \right]\sim \left\{
\begin{matrix} s^{\gamma-2},\ \ \gamma \leq 2\\
{}\\
s^0,\ \ \gamma > 2,\\
\end{matrix} \right. 
\label{m2ap}
\end{equation}

\begin{equation}
\mathcal{L}\left[ \varphi_R(t)\langle |x(t)|^3\rangle_P \right]\sim \left\{
\begin{matrix} s^{\gamma-\frac{3}{2}},\ \ \gamma \leq \frac{3}{2}\\
{}\\
s^0,\ \ \gamma > \frac{3}{2}.\\
\end{matrix} \right. 
\label{m3ap}
\end{equation}
Inserting the results (\ref{fap})-(\ref{m3ap}) into Eq. (\ref{msdap}) we find the overall MSD in the large time limit. The temporal scaling depends critically on the value of the exponent $\gamma$ as follows

\begin{equation}
\langle x^2(t)\rangle\sim \left\{
\begin{matrix} t,\ \ 0<\gamma \leq 1\\
{}\\
  t^{2-\gamma} ,\ \ 1<\gamma \leq 2   \\
{}\\
  t^0  ,\ \ \gamma >2    \\ 
\end{matrix} \right. 
\label{msdt}
\end{equation}
where, for $\gamma>2$ the following stationary value is reached:

\begin{equation}
 \langle x^2\rangle  _{s}  = \frac{  2D }{r(\gamma-2)    }\frac{1+\frac{\Gamma(\gamma -3/2)}{v\Gamma(\gamma-2)}\sqrt{rD} }{1+ \frac{1}{v}\frac{\Gamma(\gamma-1/2)}{\Gamma(\gamma-1)}\sqrt{rD}  }.
\label{msdsr}
\end{equation}

Therefore, when $0<\gamma\leq1$ the overall MSD is diffusive, when $1<\gamma<2$ is subdiffusive and when $\gamma \geq2$ there is stochastic localization, i.e. it saturates to a constant value with the consequent formation of a NESS. So that, as $\gamma$ increases, the resetting PDF decays faster, i.e., the reset rate increases by hindering the overall transport process. It is interesting to note that if the instantaneous resetting limit ($v \rightarrow \infty$) is taken, the asymptotic scaling of the MSD remains the same. This can be explained with the result in Section \ref{SecIII} in which we have found that the resetting PDF always decays slower than the returning PDF. This means that, at long times, the former becomes more relevant than the latter and therefore the effect of the latter is negligible. Seeing this, the asymptotic equivalence of the MSD scaling for instantaneous and non-instantaneous resetting is not surprising. In fact, this result resembles what has been recently found for a stochastic motion with a residence period after resetting \cite{MaCaMe19p}. There, it is shown that when both the resetting and the residence PDFs are long tailed, the residence only affects the asymptotic transport properties of the overall process when its PDF decays slower than the resetting PDF (i.e. it becomes more relevant at long times).

In Eq. (\ref{msdsr}) one can see that the stationary MSD is sensible to the returning velocity (see Figure \ref{Fig4}B for numerical confirmation) and in this case the MSD is always an increasing function of $v$. In addition, the Pareto PDF for resetting times makes possible the coexistence of a subdiffusive behavior (see the MSD in Eq. (\ref{msdt}) for $1<\gamma<2$) with the existence of a NESS given by Eq. (\ref{rspareto}). This counterintuitive phenomenon is explained by noticing that the NESS for this case has divergent MSD. The subdiffusive scaling measures then the speed at which the second moment of $\rho(x,t)$ diverges. Actually, the NESS in Eq. (\ref{rspareto}) exhibits the asymptotic behavior $\rho_s(x)\sim 1/| x|^{2\gamma -1}$ when $x^2\gg 2D/r$, which resembles the tail of a L\'evy distribution precisely when  $1<\gamma <2$.

\section{Conclusions}
\label{SecVI}

We have developed a two state model to describe resetting as a non-instantaneous movement towards the origin. In one of the states the walker is exploring and performs a random walk, while in the other it travels ballistically until it reaches the origin to start exploring again. In some way, the returning state can be seen as an exploration cost, which depends on both the type of movement in the exploring state and its duration.

We firstly focus on the case where the resetting (exploring) times are drawn from an exponential distribution and we derive an expression for the stationary distribution attained in this case. It is seen that it does not depend on the returning velocity for a diffusion process and the distribution for the exponential instantaneous resetting case is recovered \cite{EvMa11}. Regarding the stationary value of the MSD, a general formula is found in terms of the first, second and third absolute-value moments of the propagator, the resetting rate and the returning velocity. It is seen to be an increasing function of the returning velocity for exploring motions which are more likely to stay close to the origin (fBm with $\alpha < 1$) and a decreasing function when the exploring motion is more likely to occur away from it (fBm with $\alpha > 1$). Therefore, depending on the type of exploration, increasing the returning velocity may help or harm to have a bigger area of influence.

In the case of diffusion with resetting at times drawn from a Pareto PDFs, we find that the asymptotic scaling of the MSD in the non-instantaneous resetting model does not depend on the returning velocity and, consequently, is equivalent to the scaling observed in the instantaneous resetting limit. This is, the MSD scales linearly with time for $\gamma\leq 1$, when $1 < \gamma \leq 2$ the overall transport is subdifussive and for $\gamma > 2$ the MSD reaches a stationary value. In this case there is a NESS with a shape that depends explicitly on the returning velocity.

From a practical point of view, the results of this work may also help to understand the underlying dynamics of certain processes. For instance, in a central-place foraging context where animals explore their environment and occasionally return to their nest, the model gives the particular relation between the dynamics of exploration and return to the nest, on one side, and the stationary distribution on the other. This relation could in principle be validated by empirical data.

\section*{Acknowledgements}
This research has been supported by the Ministerio de Economia y Competitividad through Grant No. CGL2016-78156-C2-2-R.

\bibliography{Referencesa}
\end{document}